\documentclass[12pt]{article}
\usepackage[top = 2 cm, bottom = 3 cm, left = 3 cm, right = 2 cm]{geometry}
\usepackage{authblk, cite, color, amssymb, amsmath}
\usepackage[colorlinks = true, linkcolor = blue, citecolor = red]{hyperref}
\usepackage[dvipsnames]{xcolor}

\begin{document}

\title{Cosmological Constant in the Thermodynamic Models of Gravity}

\author{\bf Merab Gogberashvili$^{1,2}$ and Ucha Chutkerashvili$^1$}
\affil{\small $^1$ Javakhishvili Tbilisi State University, 3 Chavchavadze Avenue, Tbilisi 0179, Georgia \authorcr
$^2$ Andronikashvili Institute of Physics, 6 Tamarashvili Street, Tbilisi 0177, Georgia}

\maketitle

\begin{abstract}
Within thermodynamic models of gravity, where the universe is considered as a finite ensemble of quantum particles, cosmological constant in the Einstein's equations appears as a constant of integration. Then it can be bounded using Karolyhazy uncertainty relation applied for horizon distances, as the amount of information in principle accessible to an external observer.

\vskip 1mm

\end{abstract}
\vskip 5mm


The cosmological constant problem (that its theoretical and observed values differ by many orders of magnitudes) is one of the biggest challenges in theoretical physics \cite{Wein, Pad-lambda}. There have been several proposals to solve this discrepancy, like considering cosmological constant as a Lagrange multiplier, as a constant of integration, as a stochastic variable, anthropic interpretation, probabilistic interpretation and so on \cite{Pad-lambda}. In this article we will concentrate on thermodynamic models of gravity \cite{Gog, Pad}, where the cosmological constant arises in the Einstein's equations as a constant of integration.

Black hole thermodynamics, the Unruh effect and some other evidences suggest that the gravitation has a fundamental connection to thermodynamics (see the review \cite{Pad}). Within thermodynamic models spacetime geometry emerges from the properties of the finite unified ensemble of quantum objects (particles - spacetime atoms). In these approaches, there could be avoided the problem of apparent disappearance of gravitational energy in geometric formulations, since the system of Einstein's equations is written as a single scalar equation of the balance of gravitational and matter heat densities,
\begin{equation} \label{Termo-Ein}
\left( R_{\mu \nu} - \frac 12 g_{\mu\nu}R \right) l^\mu l^\nu = L^2_p T_{\mu \nu} l^\mu l^\nu~,
\end{equation}
in the spirit of the first law of thermodynamics. In (\ref{Termo-Ein}) $L_p$ is the reduced Planck length and $l^\nu$ is a null vector field,
\begin{equation} \label{l^2=0}
g_{\mu\nu}l^\mu l^\nu = 0~,
\end{equation}
which is the independent variable (instead of the metrics tensor $g_{\mu\nu}$) in the variational principle of the model. Note that the scalar Einstein equation contains the same information content as the usual tensor form because it is assumed that (\ref{Termo-Ein}) holds for all $l^\nu$.

The right hand side of (\ref{Termo-Ein}) can be regarded as the matter heat density, what is obvious, for example, for the case of ideal fluid using Gibbs-Duhem relation,
\begin{equation}
T_{\mu \nu} l^\mu l^\nu \to \rho + P = TS ~,
\end{equation}
where $T$ is the temperature and $S$ is the entropy density of the fluid. The gravitational heat density at the left hand side of (\ref{Termo-Ein}) is the quadratic combination of covariant derivatives of the null vector,
\begin{equation} \label{Ein}
R_{\mu \nu} l^\mu l^\nu = \left(\nabla_\nu l_\nu\right)^2 - \nabla_\mu l^\nu \nabla_\nu l^\mu + ({\rm total~derivatives}) ~.
\end{equation}

Another advantage of the thermodynamic approach based on (\ref{Termo-Ein}) is that a cosmological constant, $\Lambda$, due to (\ref{l^2=0}), can arise as an integration constant in the usual tensor form of Einstein's equations,
\begin{equation} \label{Ein}
R_{\mu \nu} - \frac 12 g_{\mu\nu}R = L^2_p \left(T_{\mu \nu} + g_{\mu\nu}\Lambda\right)~,
\end{equation}
and is not connected with the large constant vacuum energy terms in matter Lagrangians. The constant $\Lambda$ needs to be fixed using an extra physical principle and can be identified with the amount of information accessible to an eternal observer at the horizon \cite{Pad}, since we know that information is a physical entity \cite{Land}. To find the numerical value of $\Lambda$ we suggest to use the K\'arolyh\'azy uncertainty relation \cite{karo}, which bounds the ultimate precision of information on spacetime fluctuations.

It is known that, due to the Heisenberg uncertainty relation combined with general relativity, spacetime distances cannot be measured with a better accuracy than \cite{karo}
\begin{equation} \label{delta-t}
\delta x \approx L^{2/3}_p x^{1/3} ~.
\end{equation}
On the other hand, the time-energy uncertainty relation gives that the energy of the minimal spatial cell, $\delta x^3$, cannot be smaller than
\begin{equation}
E \sim \frac 1t ~.
\end{equation}
From these two estimations, it follows that for the horizon distances $x \sim 1/H$, where $H$ is the Hubble constant, the minimal energy density in the universe, what can measure an external observer, is
\begin{equation} \label{rho}
\Lambda \sim \frac{H^2}{L_p^2 } ~.
\end{equation}
This value coincides with the observed estimations for the dark energy \cite{Maz} and can be used to bound the integration constant $\Lambda$ in the Einstein's equations (\ref{Ein}).

To conclude, in this small note we suggest to identify the cosmological constant, which within the thermodynamic model of gravity appears as a constant of integration, with the amount of information accessible to an external observer. Then its value can be obtained using the K\'arolyh\'azy uncertainty relation applied to the horizon distances, which gives the observed value of the dark energy in the universe.


\end{document}